# Structure Functions in Deep Inelastic Scattering at HERA[1]

J. Blümlein (DESY), T. Doyle (Glasgow), F. Hautmann (Oregon)
M. Klein (DESY), A. Vogt (Würzburg)

## Abstract

An introduction and summary is given of the main results achieved by working group 1: Structure Functions in Deep Inelastic Scattering at HERA. The prospects were discussed of future measurements of the structure functions $F_2$, $F_L$, $xG_3$, $F_2^{Q\overline{Q}}$ and $F_\pi$ at HERA. The results represent a long term programme of experimentation with high luminosity, different lepton beam charges, proton and deuteron beams, allowing for precision measurements. The theoretical investigations focussed on QED corrections, higher order QCD corrections for different observables, resummation of small $x$ contributions, and the detailed understanding of NLO QCD evolution codes, to allow for a high precision analysis of the forthcoming deep inelastic data.

## 1  Introduction

The discussion of prospects of structure function physics has taken place for the third time in a HERA physics workshop. In 1987, the emphasis was put mainly on high $Q^2$ structure function simulations, and the determination of parton distributions [1]. In 1991, the experimental part concentrated on the reconstruction of the deeply inelastic event kinematics and Monte Carlo simulations, while the theoretical investigations began to focus on the physics of the parton densities at very low $x$ [2].

This workshop was based on the experience gained in the first four years of running at HERA, and on recent important theoretical developments, in particular the extension of next-to-leading order (NLO) calculations to more observables, and on small $x$ resummations. One of the major goals on the theoretical side was to summarize the progress towards high–precision calculations, and to point out which further theoretical investigations are most important in order to match the accuracy expected in future high-statistics structure function data from HERA.

---

[1] Summary report of Working Group 1 of the 1996 HERA Physics Workshop

On the experimental side, the scope of the studies has been extended to the charm contribution to the proton structure functions at low $x$, and to the pion structure function. The knowledge of existing HERA data allowed for much more reliable simulations than in previous workshops. Demands on the future HERA running were derived, which are contained in the experimental section of this paper. A challenging scenario of running HERA for many years has been developed. It requires the accumulation of high luminosity for both lepton beam charges and also to accelerate deuterons. This will permit to reach very high precision, i.e. an error level of only a few % for the proton stucture function $F_2$, and also accurate results for $F_L$, $F_2^c$ and $F_\pi$. On the basis of this data high precision studies of QCD in the regions of very small $x$ and of very large $Q^2$ can be carried out. In the following, we summarize the main results obtained in the theoretical and experimental investigations of the working group.

## 2 Theoretical Studies

Since the 1991 HERA Physics Workshop, various theoretical developments have taken place in the field of deep–inelastic scattering (DIS). These include the extension of QED radiative corrections to more observables, the calculation of various QCD coefficient functions and scattering cross sections to $O(\alpha_s^2)$, determinations of the 3–loop corrections of a series of moments of coefficient functions and anomalous dimensions for non–singlet as well as singlet structure functions, and the resummation of certain classes of logarithmic corrections at small $x$. During the present workshop, applications of these results have been considered to future HERA data. Comparisons have been performed of the experimental requirements and the theoretical accuracy achieved at present.

### 2.1 QED Radiative Corrections

The understanding of the $O(\alpha)$ QED corrections for a few choices of kinematical variables was already obtained at the time of the 1991 HERA workshop [3]. In a contribution to the present workshop [4], recent developments in the semianalytical approach by the HECTOR collaboration [5] have been summarized. These include the $O(\alpha)$ corrections for a wide variety of kinematical variables as well as leading–log $O(\alpha^2)$ results. Recently also the QED corrections due to the resummation of the $O(\alpha[\alpha \ln^2(1/x)]^l \log[Q^2/m_e^2])$ terms were calculated [6]. For leptonic variables, their contribution is of similar size as the second order corrections $\sim \alpha^2 \log(Q^2/m_e^2)$ and diminishes the effect of these terms. A first complete $O(\alpha)$ calculation was reported also for neutral current polarized–lepton polarized–hadron scattering [7] accounting for photon and $Z$-boson exchange. This calculation covers the cases of both longitudinal and transverse nucleon polarization for the twist–2 contributions to the Born term [8].

### 2.2 Comparisons of NLO QCD Evolution Codes and the Theoretical Error of $\alpha_s(M_Z^2)$

A major effort within the working group concerned a detailed comparison of next-to-leading order (NLO) QCD evolution codes. The aim of this comparison was to understand the accuracy of the different numerical solutions of the evolution equations and their conceptual differences. A numerical agreement of better than $\pm 0.05\%$ of the parton densities has been achieved between five of the evolution programs. For these codes, previous deviations due to different theoretical assumptions on the truncation of the perturbative series at NLO are now completely understood. The results of this study are summarized in ref. [9].

Precision measurements of $F_2$ allow for detailed QCD tests and an improved extraction of $\alpha_s(Q^2)$. The theoretical uncertainties of the $\alpha_s$ measurement from scaling violations have been investigated for NLO analyses in the HERA range [10, 11]. The different representations of the evolution within NLO lead to a shift of up to $\Delta\alpha_s(M_Z^2) = 0.003$. The largest theoretical errors at NLO are due to the renormalization ($R$) and factorization ($M$) scale uncertainties, resulting in $\Delta\alpha_s(M_Z^2) = {}^{+0.004}_{-0.006}|_R {}^{-0.003}_{+0.003}|_M$, for a $Q^2$–cut of $Q^2 \geq 50\,\text{GeV}^2$. The contribution of mass threshold uncertainties to $\Delta\alpha_s(M_Z^2)$ was conservatively estimated to 0.002 in ref. [12]. Due to the large statistics at low $Q^2$ it appears to be desirable to include also the range down to a few GeV$^2$ in the QCD analyses. To exploit this region fully, however, requires to carry out next-to-next-to leading (NNLO) analyses, for which the 3–loop splitting functions still have to be calculated.

## 2.3 $O(\alpha_s^2)$ and $O(\alpha_s^3)$ Corrections

Since the 1991 HERA Physics workshop, various two–loop calculations have been performed for quantities related to structure functions. A survey on the status of these calculations, and recent three–loop results, was presented to this working group [13]. The NLO corrections turn out to be essential for the quantitative understanding of most of the observables.

Numerical studies during this workshop were devoted to the behaviour of $F_L(x, Q^2)$ [14] and $F_2^{Q\overline{Q}}$ [15, 16]. An update of the parametrization of the NLO heavy flavour coefficient functions is also given in [15]. Moreover, a calculation of the heavy flavour structure functions in charged current interactions has been presented [17]. Besides of the twist-2 contributions to the structure function $F_L$, also higher–twist terms for its non–singlet part have been investigated, accounting for renormalon contributions [18]. Also the NLO corrections to the scattering cross section $\sigma(\gamma + g \to J/\psi + X)$ are available now [19].

First results on the behaviour of structure functions in NNLO were reported by the NIKHEF group [20, 21]. The first moments for both the non–singlet and singlet combinations of unpolarized DIS structure functions have been calculated. By the same group also a phenomenological analysis was presented [22], estimating the $x$-dependence of the corresponding splitting functions by a fit allowing for a set of functions which are known to contribute. Also a phenomenological application of the non–singlet results to $xF_3$ was reported [23].

## 2.4 Resummations for Small $x$

The measurement of DIS structure functions remains one of the major methods to investigate the small-$x$ physics at HERA. Various aspects have been considered in the working group. A possible approach has been discussed which relies on the $k_\perp$-factorization method [24]. It consists of combining systematically the resummation of the small-$x$ logarithmic corrections, as given by the BFKL formalism [25] and the QCD corrections to it, with the QCD mass factorization theorem, dictated by the renormalization group. This approach enables one to study the small-$x$ effects by solving improved evolution equations which include resummed kernels. A first numerical analysis along these lines was carried out in ref. [26]. A review of these equations and the current status of resummed calculations, covering also the non–singlet cases, can be found in ref. [27].

During the workshop numerical studies of structure functions at small $x$ have been performed by two groups [28, 29]. One contribution to the workshop [30] dealt with the resummation of the small-$x$ contributions on the level of a double–log approximation. In the flavour non–singlet sector resummation effects are small, less than 1% [31]. In the singlet sector, however, they may give rise to large contributions [28, 29]. The question of assessing the importance of unknown sub–leading terms has also been addressed in ref. [28], comparing several different models. The outcome is that these terms seem to be able to affect the result sizeably. This indicates that at present the uncertainties on the theoretical predictions at small-$x$ are fairly large, and more accurate calculations (next-to-leading small-$x$ logarithms as well as exact 3-loop contributions) are necessary. The analysis in ref. [29] has emphasized the role of a combined determination of $F_2$ and $F_L$ to pin down the behaviour of the QCD perturbative series at small $x$. The effects of small-$x$ resummations on the photon structure functions were also discussed [28].

On the leading–order level, predictions for the structure functions, covering the BFKL effects at small $x$, can also be obtained starting from a different equation, which was introduced a few years ago [32] to describe the detailed structure of the gluon radiation associated to small-$x$ events. In ref. [33], the solution to this equation has been investigated numerically for the structure function $F_2$.

In the working group also the properties of the BFKL resummation equation itself were discussed with emphasis on the transverse momentum cut–off [34]. Recent progress in calculating NLO corrections to the BFKL kernel was reported in ref. [35], see also [27] for other ongoing investigations.

An alternative formulation of QCD at small $x$ has been proposed in a series of papers, based on a colour dipole concept [36]. The theory has been worked out for the case of scattering of two quarkonia, in which non-perturbative effects are suppressed by the smallness of the quarkonium radius. First attempts to apply the colour dipole formulation to deep–inelastic scattering have been reported at this workshop [37, 38]. The suitability of this approach to investigate unitarity corrections and parton saturation has been emphasized, and an explicit parametrization of multi-pomeron exchange contributions has been presented [38].

# 3 Experimental Studies

The future HERA measurements of structure functions using electron and positron, proton and deuteron beams promise to be of great interest, since the envisaged increase of luminosity will allow for reducing the statistical and the systematic errors, especially for the proton case, to the level of a few per cent in almost the full accessible kinematic range. Within this working group, detailed simulation studies have been performed of various structure function measurements, in order to estimate their expected accuracy and to analyze their physics impact.

## 3.1 The Proton Structure Function $F_2$

A thorough simulation has been carried out [39] of future HERA measurements of the proton structure function $F_2(x, Q^2)$ for a nominal kinematic range given by $y < 0.8$, $\theta_e < 177^o$, $Q^2 > 0.5$ GeV$^2$, and $\theta_h > 8^o$. The following assumptions were made on the future measurement accuracies: 0.5-1% for the scattered electron energy $E'_e$, 0.5-1 mrad for the polar angle $\theta_e$, 2% on the hadronic energy $E_h$, 1-2% for the photoproduction background uncertainty at high $y$, and 2% due to trigger and detector efficiencies. The luminosity was assumed to be known within 1%. Moreover, control of the radiative corrections at the level of 1% has been assumed. These conditions lead to an estimated systematic error of $F_2$ of about 3% in almost the full kinematic range of $2 \cdot 10^{-5} \leq x \leq 0.7$ and $0.5 \leq Q^2 \leq 5 \cdot 10^4$ GeV$^2$. The anticipated accuracy represents an improvement by about a factor of two compared to the present results but extended to a much wider region. The current H1 and ZEUS structure function analyses served as a basis for the simulations and were summarized in various talks presented to this working group [40].

In a detailed NLO QCD analysis [39], possible determinations of the strong coupling constant $\alpha_s$ and of the gluon distribution have been considered using the H1 and the ZEUS QCD analysis programs and fitting techniques. The error of $\alpha_s$ strongly depends on the minimum $Q^2$ which can be included into such analysis. While perturbative QCD seems to work down to 1 GeV$^2$ at low $x$, the theoretical scale uncertainties in NLO become very large, see above. With only HERA proton data an $\alpha_s$ error of about 0.004 can be expected for $Q^2 > 3$ GeV$^2$. As described in [39] the measurement of $\alpha_s$ requires very accurate control of the dependence of the systematic errors on the kinematic variables. This will permit to largely reduce their effect in the fit procedure like it has been practiced in the BCDMS/SLAC data analysis [41] which lead to an experimental $\alpha_s$ error of 0.003. Combination of the high $x$ fixed target $F_2$ data with the low $x$ HERA data promises to yield a precision measurement of $\alpha_s$ with an estimated experimental error of 0.0013 for $\alpha_s(M_Z^2)$. Simultaneously the gluon distribution can be measured very accurately with an estimated error of e.g. 3% for $x = 10^{-4}$ and $Q^2 = 20$ GeV$^2$ [39] using HERA data only. This potential measurement can only be reliably interpreted if the theoretical description is extended to NNLO.

## 3.2 High $Q^2$ Structure Functions

For $Q^2 \gtrsim 500$ GeV$^2$ and with both electron and positrons beams employed, there are two neutral current and two charged current reactions, i.e. four cross sections to be measured. This permits to extract various structure functions and combinations of parton densities, as was already thoroughly studied in the previous HERA workshops [42, 43] and summarized in the discussion [44]. The main conclusions remain valid: i) the charge asymmetry in neutral current $e^\pm p$ scattering allows to determine the interference structure function $xG_3 = 2x \sum Q_q a_q \cdot (q - \bar{q})$ for large $y$; ii) the charged current cross sections directly determine the valence quark distributions for $x \geq 0.3$; iii) with a high precision measurement of the four cross sections, various combinations of parton distributions can be unfolded, e.g. the singlet distribution or the strange quark density. All these measurements require highest luminosities, $\mathcal{L} \geq 500$ $pb^{-1}$, about equally shared between electron and positron runs.

## 3.3 Deuteron Structure Functions

Interest has been expressed in the structure function working group [39, 45] in measuring the electron–deuteron DIS cross section at HERA. At low $x$, despite a few % structure function extraction uncertainty due to shadowing effects [46], the proton–deuteron $F_2$ difference will permit to constrain the up-down quark difference. This is expected to vanish towards low $x$ but has not been measured yet in the domain of $x < 10^{-3}$. Deuteron structure function data are important for a self consistent QCD analysis of HERA data as they constrain the non–singlet distributions in the $\alpha_s$ and gluon determination. Even with a luminosity of $\mathcal{L} = 50\text{pb}^{-1}$ only, interesting parton distribution combinations as $s - c$ [43] at high $Q^2 > 100$ GeV$^2$ can be measured.

## 3.4 Changing the Beam Energies

With the maximum possible energies of $E_p \simeq 1000$ GeV and $E_e \simeq 35$ GeV, the cms. energy squared can be increased by about a factor of 1.5 as compared to the nominal energy runs. Hence $x$ values lower by this factor can be accessed at a given $Q^2$. It should be stressed that this high-energy option is the only way to explore this extended kinematical range in the foreseeable future. It requires only modest luminosity $\mathcal{L} \simeq 5$ pb$^{-1}$ for the exploration of the low $x$ region.

Because of the relation $Q^2 \simeq (2E_e \cot \theta_e/2)^2$, the decrease of the electron beam energy can be employed to reach very low values of $Q^2 < 1$ GeV$^2$. Using a small luminosity, $\mathcal{L} \simeq 1$ pb$^{-1}$, only this will allow for a high precision measurement of the transition region to photoproduction even after the foreseen luminosity upgrade which may limit the detector acceptance close to the beam pipe.

A run at lowest possible proton beam energy with $\mathcal{L} \simeq 10$ pb$^{-1}$ is required in order to ensure maximum overlap of the HERA $F_2$ data with the fixed–target results. This is important for the precision measurement of $\alpha_s$. Lowering $E_p$ is also a necessity for measuring the longitudinal structure function $F_L$.

## 3.5 The Longitudinal Structure Function $F_L$

Various, partially complementary studies were presented to the workshop on the estimated measurement accuracy of the longitudinal structure function $F_L$ [47, 48]. The joint conclusion of these studies is that an $F_L$ precision measurement requires a set of about four different proton beam energies, $E_p = 250, 350, 450$ and $820$ GeV, for instance, with luminosities around 5–10 $pb^{-1}$ per energy setting. In a wide range of $Q^2$, between about 4 and 100 GeV$^2$, $F_L$ should be measurable with an absolute error of typically 0.08 [48]. That accuracy largely depends on the maximum $y$, i.e. the minimum electron energy which can be controled in view of background and trigger requirements. The availability of various proton energies and the use of the subtraction method [49] will permit a significant measurement of the $x$ dependence of $F_L(x, Q^2)$ which is important to distinguish between different predictions for $F_L$. A run at highest proton energy, $E_p \simeq 1000$ GeV, would represent an important gain in sensitivity to $F_L$, and lower $E_e$ data would provide useful systematics cross checks.

## 3.6 The Charm Contribution to $F_2$

At low $x$, the charm contribution to the inclusive DIS cross section has been measured to be about one quarter [50]. Its understanding is crucial for the interpretation of $F_2$ and interesting for an independent measurement of the gluon distribution. With a luminosity of several 100 pb$^{-1}$, the charm structure function will be measurable with an estimated accuracy of 10%, the error being mainly due to the limited knowledge of the charm fragmentation probability $P(c \to D)$ into D mesons, and the detector and analyses uncertainties [51, 15]. This will permit a detailed investigation of the charm production mechanism as a function of $Q^2$ and $x$ and a complementary determination of the gluon distribution. The measurement relies on the observation of open charm production at HERA which should profit from Silicon detectors installed near the interaction vertex in order to enhance the charm tagging efficiency.

## 3.7 Structure of the Pion

HERA provides an interesting opportunity to study the structure of the pion [52, 53] as was presented to this working group [54]. Finite values of momentum transfer squared from the incoming proton beam to the outgoing neutron can be measured in the H1 and ZEUS neutron calorimeters with mean values of the order of 0.1 GeV$^2$. Studies were performed prior to the workshop [53], and a complete Monte Carlo simulation of the performance of the H1 forward neutron calorimeter is described in [55]. A modest luminosity of 10 pb$^{-1}$ would yield about 5000 events in the range $10 < Q^2 < 15$ GeV$^2$. This would enable the structure of the virtual pion to be determined as a function of the longitudinal momentum of the exchanged pion for $0.6 > x_\pi > 10^{-4}$. Higher luminosities may allow the separation of the longitudinal structure of the pion and the determination of the structure of higher mass resonances.